\documentclass{article}
\usepackage{graphicx}
\usepackage{amsmath}
\usepackage{amsfonts}
\usepackage{amssymb}

\begin{document}

\title{Non-Holonomic Control III : Coherence Protection by the Quantum Zeno Effect and Non-Holonomic Control}
\author{E. Brion\\\emph{Laboratoire Aim\'{e} Cotton, }\\\emph{CNRS II, B\^{a}timent 505, }\\\emph{91405 Orsay Cedex, France.}
\and V.M. Akulin\\\emph{Laboratoire Aim\'{e} Cotton, }\\\emph{CNRS II, B\^{a}timent 505, }\\\emph{91405 Orsay Cedex, France.}
\and D. Comparat\\\emph{Laboratoire Aim\'{e} Cotton, }\\\emph{CNRS II, B\^{a}timent 505, }\\\emph{91405 Orsay Cedex, France.}
\and I. Dumer\\\emph{College of Engineering, }\\\emph{University of California, }\\\emph{Riverside, CA 92521, USA. }
\and V. Gershkovich\\\emph{Institut des Hautes Etudes Scientifiques,}\\\emph{ Bures-sur-Yvette, France. }
\and G. Harel\\\emph{Department of Computing, }\\\emph{University of Bradford, }\\\emph{Bradford, West Yorkshire BD7 1DP, United Kingdom. }
\and G. Kurizki\\\emph{Department of Chemical Physics, }\\\emph{Weizmann Institute of Science, }\\\emph{76100 Rehovot, Israel. }
\and I. Mazets\\\emph{Department of Chemical Physics, }\\\emph{Weizmann Institute of Science, }\\\emph{76100 Rehovot, Israel. }\\\emph{A.F. Ioffe Physico-Technical Institute, }\\\emph{194021 St. Petersburg, Russia. }
\and P. Pillet\\\emph{Laboratoire Aim\'{e} Cotton, }\\\emph{CNRS II, B\^{a}timent 505, }\\\emph{91405 Orsay Cedex, France.}}
\maketitle
\begin{abstract}
In this paper, we present a coherence protection method based upon a multidimensional generalization of the Quantum Zeno Effect, as well as ideas from the coding theory. The non-holonomic control technique is employed as a physical tool which allows its effective implementation. The two limiting cases of small and large quantum systems are considered.   
\end{abstract}

\section{Introduction}

The uncontrollable interaction of a quantum system with its environment is responsible for \textquotedblright
quantum errors\textquotedblright which lead to a partial or complete loss of the information initially stored in its quantum state. After Shor's demonstration \cite{Sho95} that error-correcting schemes exist in quantum computation, a general framework of error-correction has been built
upon the formalism of quantum operations. The main contributions concern
quantum codes, and particularly the class of stabilizer codes ; other
strategies developed suggest the use of \textquotedblright noiseless quantum
codes\textquotedblright\ or \textquotedblright decoherence-free
subspaces\textquotedblright. All these methods usually demand that errors act
independently on different qubits (the independent error model), and make use
of the symmetry properties associated with these requirements. This implies
that the set of errors to be corrected is restricted to a special subgroup,
called the Clifford group. In this paper, we present a protection method which does not make so drastic assumptions.

For low dimensional systems, one can take advantage of the Quantum Zeno Effect
allied with basic ideas of coding theory in order to protect the information
encoded on a subspace of the total state space. To be more explicit, one
frequently repeats a three step sequence comprising coding, decoding and
projection, which prevents errors from developing in the system : coding and
decoding consist in the application of a unitary matrix and its inverse, which
can be achieved through non-holonomic control, and act in such a way that
erroneous infinitesimal components are orthogonal to the information subspace
; projection is performed through an irreversible process such as spontaneous
emission which clears the unwanted orthogonal increments out.

Promising for small quantum systems, this method becomes however exponentially
complex when the number of the qubits involved increases. For large quantum
systems, we suggest to employ random coding to reduce the influence of errors
consisting in binary interactions. In this context, high dimensionality does
not appear as an impediment, but rather as an advantage, since it 'dilutes'
the influence of the errors.

This paper is organized as follows. In the second section, we show that the Quantum Zeno Effect allied with basic ideas of coding theory allows one to protect the quantum information contained in
low-dimensional quantum systems. In the third section, we present the random coding technique which can
protect the information stored in large systems against the errors resulting
from binary interactions.

\section{Coherence Protection in low-dimensional systems : Quantum Zeno Effect and Non-Holonomic Control}

The Quantum Zeno Effect (QZE) \cite{MS77, KK00} appears in a system which is frequently measured in its (necessarily known) initial state: if the time interval between two projective measurements is small enough, the evolution of the system is nearly ''frozen'' ; in other words the one-dimensional subspace spanned by the (necessarily known) initial state is protected against the influence of the natural Hamiltonian of the system. We suggest to generalize this effect in order to protect any (unknown) vector of a given multidimensional subspace of the whole Hilbert space. To this end, we propose an information protection scheme, described in the first paragraph of this section \cite{Brion04}, as well as the algorithmic tools which allow its implementation, and which are presented in the second paragraph. 

\subsection{Multidimensional Zeno Effect and Coherence Protection\label{II}}

In this paragraph, we shall first present the multidimensional QZE which allows us to protect an arbitrary subspace of the Hilbert space against the action of a set of given interaction Hamiltonians. Then, we shall take advantage of this phenomenon to protect an information-carrying subsystem of a compound quantum system from the influence of some uncontrolled error-inducing external fields.

\bigskip

Consider a quantum system $\mathcal{S}$, whose $N$-dimensional Hilbert space
is denoted by $\mathcal{H}$ and whose time-dependent Hamiltonian has the form
\begin{equation}
\widehat{H}(\tau)=\sum_{m=1}^{M}f_{m}(\tau)\widehat{E}_{m},
\label{Hamiltonian}%
\end{equation}
where $\left\{  \widehat{E}_{m}\right\}  _{m=1,...,M}$ are $M$ given
independent Hermitian matrices on $\mathcal{H}$ and $\left\{  f_{m}%
(\tau)\right\}  _{m=1,...,M}$\ are $M$ unknown functions of time. The
Hamiltonian $\widehat{H}(\tau)$ accounts for the errors we want to get rid of.
Note that the unperturbed part of the Hamiltonian (\ref{Hamiltonian}) is
assumed to be zero (or proportional to the identity so that one can set it to
zero). The standard QZE allows us to nearly ''freeze'' the evolution of the system by measuring it frequently
enough in its (known) initial state ; in other words, through this effect we can protect the one-dimensional subspace spanned by the initial state of the system from the influence of the error-inducing Hamiltonian (\ref{Hamiltonian}). In what follows, we generalize this effect so as to protect an arbitrary
multidimensional subspace $\mathcal{C}$ from $\widehat{H}(\tau)$.

Any vector $\left|  \psi\right\rangle $ of $\mathcal{C}$ evolves according to $\widehat{U}(t,t_{0})=\mathcal{T}\left\{  \exp\left[  -i\int_{t_{0}}^{t}\widehat{H}(\tau)d\tau\right]  \right\}  ,$
where $\mathcal{T}$ denotes time-ordering, and where we set $\hbar=1$. For the QZE to hold, we shall only consider evolution in short time periods, the duration $\tau_{Z}$ of which is so short that the corresponding action of the $M$ components
of the Hamiltonian (\ref{Hamiltonian}) is small, \emph{i.e.} $\left|  \widehat{E}_{m}\int_{t}^{t+T}f_{m}(\tau)d\tau\right|  \ll1.$ We can thus expand
\begin{equation}
\widehat{U}(t+\tau_{Z},t)=\widehat{U}_{inf}\simeq\widehat{I}-i\sum_{m=1}^{M}\left(
\int_{t}^{t+\tau_{Z}}f_{m}(\tau)d\tau\right)  \widehat{E}_{m}.\label{evolinf}%
\end{equation}
After a Zeno interval $\tau_{Z}$, the initial state $\left|
\psi\right\rangle $ is thus transformed into $\left|  \psi_{e}\right\rangle
=\left|  \psi\right\rangle +\left|  \delta\psi_{e}\right\rangle $ where
$\left|  \delta\psi_{e}\right\rangle \simeq-i\sum_{m=1}^{M}\varepsilon
_{m}\widehat{E}_{m}\left|  \psi\right\rangle $ with $\varepsilon_{m}=\left(
\int f_{m}(\tau)d\tau\right).$

\begin{figure}[ptb]
\begin{center}
\includegraphics[
height=6.5363in,
width=4.5956in
]{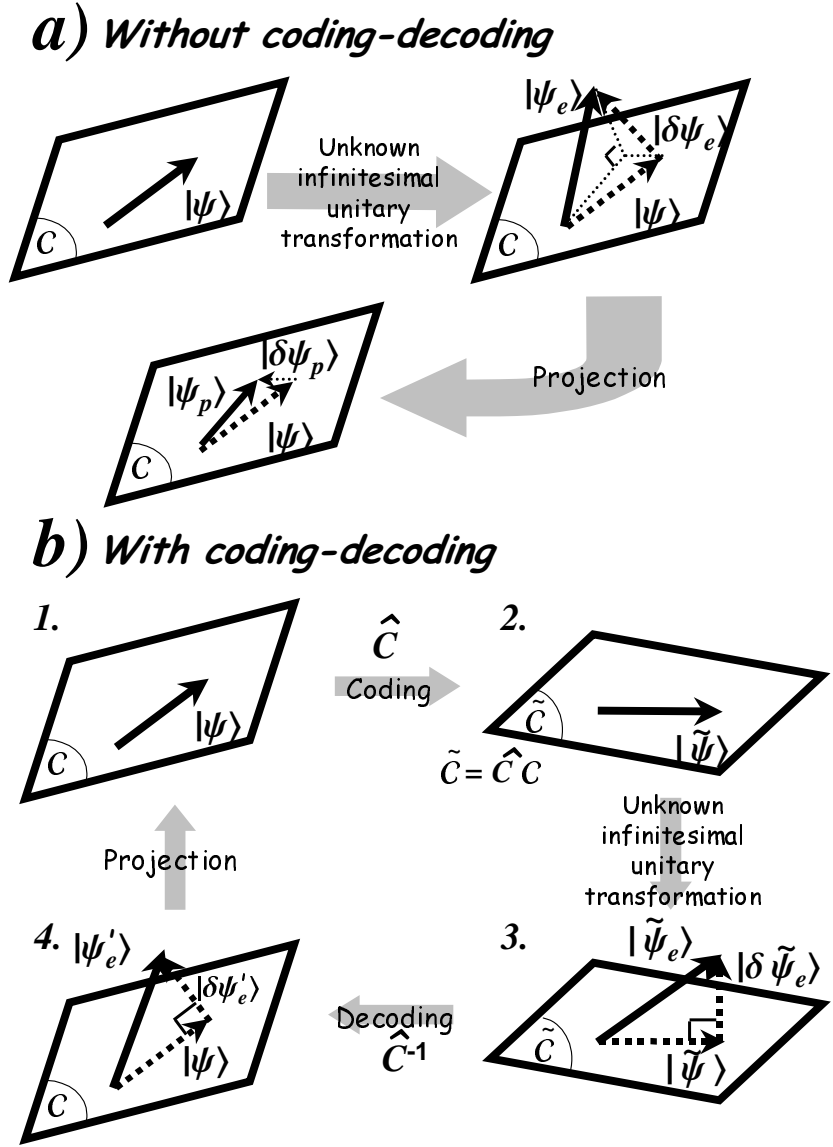}
\end{center}
\caption{Multidimensional QZE: a) a simple projection fails to recover the
initial vector, b) the sequence coding-decoding-projection protects the
initial vector.}%
\label{Fig7}%
\end{figure}

Let us assume that we are physically able to perform the measurement-induced
projection onto $\mathcal{C}$ in the system $\mathcal{S}$ (see below the
discussion of such projections for compound systems comprising an information
subsystem and an ancilla). If we straightforwardly apply the standard QZE procedure by merely projecting the state vector $\left|  \psi_{e}\right\rangle ,$
resulting from the infinitesimal evolution of the initial state $\left|
\psi\right\rangle ,$ onto $\mathcal{C}$, we get the vector $\left|  \psi
_{p}\right\rangle $, which, a priori, differs from $\left|  \psi\right\rangle
$ (see Fig.\ref{Fig7}a), since, usually, the vectors $\widehat{E}_{m}\left|
\psi\right\rangle $\ and thus the increment vector $\left|  \delta\psi
_{e}\right\rangle $\ itself are not orthogonal to $\mathcal{C}$.  
It is thus clear that we have to adapt the standard Zeno strategy.

To this end, we assume a unitary matrix $\widehat{C}$ acting on $\mathcal{H}$,
which we call the coding matrix, such that the Hermitian operators $\left\{
\widehat{E}_{m}\right\}  _{m=1,...,M}$ act orthogonally on the subspace
$\widetilde{\mathcal{C}}=\widehat{C}\mathcal{C}$, called the code
space. Let us denote by $I\geq1$ the dimension of $\mathcal{C}$ and by
$\left\{  \left|  \gamma_{i}\right\rangle \right\}  _{i=1,...,I}$ one of its
orthonormal bases ; $\left\{  \left|  \widetilde{\gamma}_{i}\right\rangle
=\widehat{C}\left|  \gamma_{i}\right\rangle \right\}  _{i=1,...,I}$ will
denote one of the orthonormal bases of $\widetilde{\mathcal{C}}$, the state
vectors $\left|  \widetilde{\gamma}_{i}\right\rangle $ being called the
codewords. For any pair $\left(  \left|  \widetilde{\gamma}_{s}\right\rangle
,\left|  \widetilde{\gamma}_{t}\right\rangle \right)  $ of codewords and any
operator $\widehat{E}_{m}\in\left\{  \widehat{E}_{m}\right\}  _{m=1,...,M}$ we
have, by the definitions of $\widehat{C}$ and $\widetilde{\mathcal{C}}$
\begin{align}
\left\langle \widetilde{\gamma}_{t}|\widetilde{\gamma}_{s}\right\rangle  &
=\delta_{st}\text{ \ \ (orthonormality conditions)}\label{bascod1}\\
\left\langle \widetilde{\gamma}_{t}\left|  \widehat{E}_{m}\right|
\widetilde{\gamma}_{s}\right\rangle  &  =0\text{ \ \ (orthogonality of the
errors)}\label{bascod2}%
\end{align}
Equivalently, for any pair $\left(  \left|  \psi\right\rangle ,\left|
\chi\right\rangle \right)  $ of vectors of $\mathcal{C}$ and for any operator
$\widehat{E}_{m}\in\left\{  \widehat{E}_{m}\right\}  _{m=1,...,M}$
\begin{equation}
\left\langle \chi\left|  \widehat{C}^{\dagger}\widehat{E}_{m}\widehat
{C}\right|  \psi\right\rangle =0.\label{matcod}%
\end{equation}
In particular, for any pair $\left(  \left|  \gamma_{s}\right\rangle ,\left|
\gamma_{t}\right\rangle \right)  $ of basis vectors of $\mathcal{C}$ and for
any operator $\widehat{E}_{m}\in\left\{  \widehat{E}_{m}\right\}
_{m=1,...,M}$
\begin{equation}
\left\langle \gamma_{t}\left|  \widehat{C}^{\dagger}\widehat{E}_{m}\widehat
{C}\right|  \gamma_{s}\right\rangle =0.\label{matcodbis}
\end{equation}
If we apply the coding matrix $\widehat{C}$ to the initial state vector
$\left|  \psi\right\rangle $, before exposing it to the action of the
Hamiltonian (\ref{Hamiltonian}), we obtain the new vector $\left|
\widetilde{\psi}\right\rangle =\widehat{C}\left|  \psi\right\rangle
\in\widetilde{\mathcal{C}}$ (Fig.\ref{Fig7}b1,2) which is transformed after a
Zeno interval $\tau_{Z}$ into $\left|  \widetilde{\psi}_{e}\right\rangle =\widehat
{U}_{inf}\left|  \widetilde{\psi}\right\rangle =\left|  \widetilde{\psi
}\right\rangle +\left|  \delta\widetilde{\psi}_{e}\right\rangle ,$ where
$\left|  \delta\widetilde{\psi}_{e}\right\rangle \simeq-i\sum_{m=1}%
^{M}\varepsilon_{m}\widehat{E}_{m}\left|  \widetilde{\psi}\right\rangle
=-i\sum_{m=1}^{M}\varepsilon_{m}\widehat{E}_{m}\widehat{C}\left|
\psi\right\rangle $ (Fig.\ref{Fig7}b3).\ Decoding $\left|  \widetilde{\psi
}_{e}\right\rangle $\ yields the vector $\left|  \psi_{e}^{\prime
}\right\rangle =\widehat{C}^{-1}\left|  \widetilde{\psi}_{e}\right\rangle
=\left|  \psi\right\rangle +\left|  \delta\psi_{e}^{\prime}\right\rangle $
where $\left|  \delta\psi_{e}^{\prime}\right\rangle \simeq-i\sum_{m=1}%
^{M}\varepsilon_{m}\widehat{C}^{\dagger}\widehat{E}_{m}\widehat{C}\left|
\psi\right\rangle $. From Eq.(\ref{matcod}) it can be seen that for any vector
$\left|  \chi\right\rangle \in\mathcal{C}$, $\left\langle \chi|\delta\psi
_{e}^{\prime}\right\rangle =-i\sum_{m=1}^{M}\varepsilon_{m}\left\langle
\chi\right|  \widehat{C}^{\dagger}\widehat{E}_{m}\widehat{C}\left|
\psi\right\rangle =0$ which means that $\left|  \delta\psi_{e}^{\prime
}\right\rangle $ is orthogonal to $\mathcal{C}$ (Fig.\ref{Fig7}b4). A
measurement-induced projection onto $\mathcal{C}$\ finally recovers the
initial vector $\left|  \psi\right\rangle $ with a probability very close to
$1$ (the error probability is proportional to $\tau_{Z}^{2}$). If the coding-decoding-projection sequence is frequently repeated, any vector $\left|  \psi\right\rangle $ of the subspace $\mathcal{C}$\ can thus be protected from the Hamiltonian (\ref{Hamiltonian}) for as long as needed.

The multidimensional generalization of the QZE we have just presented allows one to protect any subspace $\mathcal{C}$ of a Hilbert space $\mathcal{H}$ against Hamiltonians of the form (\ref{Hamiltonian}), and is thus very useful in the context of information protection as we shall see in the following. 

\bigskip

Indeed, let us consider an information system $\mathcal{I}$ of Hilbert space $\mathcal{H}_{I}$ and dimensionality $I$. This system is subject to a set of $M$
error-inducing Hamiltonians $\left\{  \widehat{E}_{m}\right\}  _{m=1,...,M}$
which, for instance, represent the interactions of the system with $M$ uncontrolled external classical fields $f_{m}(t)$: we want to get rid of this external influence which is likely to result in the loss of the information stored in the initial state vector $\left|  \psi_{I}\right\rangle =\sum_{i=1}^{I}c_{i}\left|  \nu_{i}\right\rangle $, where $\left\{  \left|  \nu
_{i}\right\rangle \right\}  _{i=1,...,I}$ denotes an orthonormal basis of
$\mathcal{H}_{I}$. To this end, we shall use the multidimensional Zeno Effect. As the multidimensional QZE can only protect a subspace of the whole Hilbert
space, we first have to add an $A$-dimensional auxiliary system $\mathcal{A}$ (called ancilla) to our system $\mathcal{I}$, so that the information is
transferred from $\mathcal{H}_{I}$\ into an $I$-dimensional subspace
$\mathcal{C}$ of the $\left(  N=I\times A\right)  $-dimensional Hilbert space
$\mathcal{H}=\mathcal{H}_{I}\mathcal{\otimes H}_{A}$\ of the compound system $\mathcal{S=I\otimes A}$. Furthermore, we shall suppose that all the state vectors of the different Hilbert spaces $\mathcal{H}_{I}$, $\mathcal{H}_{A}$ and hence $\mathcal{H}$\ are degenerate in energy so that the unperturbed part $\widehat{H}_{0}$ of the Hamiltonian can be set to zero as in the first part
of this section: the subspace $\mathcal{C}$ and the information it carries can
thus be protected through the multidimensional QZE. Note that $\mathcal{A}$ and
$\mathcal{I}$ need not be ''physically separate'' systems, but only have to
possess independent Hilbert spaces $\mathcal{H}_{A}$ and $\mathcal{H}_{I}$.

Let us now return to our problem and first consider the simple case when the ancilla is initially in the pure state $\left|  \alpha\right\rangle $. The information initially stored by $\left|  \psi_{I}\right\rangle
\in\mathcal{H}_{I}$ is transferred into the factorized state $\left|
\psi\right\rangle =\left|  \psi_{I}\right\rangle \otimes\left|  \alpha
\right\rangle =$ $\sum_{i=1}^{I}c_{i}\left|  \nu_{i}\right\rangle
\otimes\left|  \alpha\right\rangle =$ $\sum_{i=1}^{I}c_{i}\left|  \gamma
_{i}\right\rangle $ of
$\mathcal{C}=\mathcal{H}_{I}\mathcal{\otimes}Span\left[  \left|
\alpha\right\rangle \right]  =Span\left[  \left\{  \left|  \gamma
_{i}\right\rangle =\left|  \nu_{i}\right\rangle \otimes\left|  \alpha
\right\rangle \right\}  _{i=1,...,I}\right]  $. Equivalently, the initial density matrix of the compound system $\mathcal{S}$ is $\widehat{\rho}=\left(
\left|  \psi_{I}\right\rangle \left\langle \psi_{I}\right|  \right)
\otimes\left(  \left|  \alpha\right\rangle \left\langle \alpha\right|
\right)$, which is transformed after the coding step into
$\widehat{\widetilde{\rho}}=\widehat{C}^{\dagger}\widehat{\rho}\widehat{C}$ ;
at the end of the action of the errors it is transformed into $\widehat
{\widetilde{\rho}}_{e}=\widehat{U}_{inf}^{\dagger}\widehat{C}^{\dagger
}\widehat{\rho}\widehat{C}\widehat{U}_{inf}$ ; finally,
after decoding, it takes the form
$\widehat{\rho}_{e}=\widehat{C}\widehat{U}_{inf}^{\dagger}\widehat{C}^{\dagger}\widehat{\rho}\widehat{C}\widehat{U}_{inf}\widehat{C}^{\dagger}$. In this setting, the projection onto $\mathcal{C}$ can be
simply achieved by measuring the ancilla in its initial state $\left|
\alpha\right\rangle $. As $\tau_{Z}$ is very short, the state of the ancilla evolves just a little within a Zeno interval : the probability of detecting it in its initial state $\left|  \alpha\right\rangle$, and thus of projecting
the state of the compound system onto $\mathcal{C}$ is thus very close to $1$.
After projection, we trace out the ancilla to obtain the final reduced density matrix $\widehat{\rho}_{I}^{\prime}=\left\langle \alpha\right|  \widehat
{C}\widehat{U}_{inf}^{\dagger}\widehat{C}^{\dagger}\widehat{\rho}\widehat
{C}\widehat{U}_{inf}\widehat{C}^{\dagger}\left|  \alpha\right\rangle $ for the
information system $\mathcal{I}$; in the same way, one can calculate the initial reduced density matrix is $\widehat{\rho}_{I}=\left|  \psi
_{I}\right\rangle \left\langle \psi_{I}\right|  .$ The variation
$\delta\widehat{\rho}_{I}=$ $\widehat{\rho}_{I}^{\prime}-\widehat{\rho}_{I}$
of the information-space density matrix during the whole process can then be expressed as the commutator
\[
\delta\widehat{\rho}_{I}=-i\left[  \sum_{m=1}^{M}\int f_{m}(\tau
)d\tau\left\langle \alpha\right|  \widehat{C}^{\dagger}\widehat{E}_{m}%
\widehat{C}\left|  \alpha\right\rangle ,\widehat{\rho}_{I}\right]  ,
\]
from which we infer that $\widehat{\rho}_{I}$ satisfies the equation $
i\frac{d\widehat{\rho}_{I}}{dt}  = \left[  \widehat{h}_{e},\widehat{\rho
}_{I}\right],$ where $\widehat{h}_{e}  = \sum_{m=1}^{M}f_{m}\left\langle \alpha\right|
\widehat{C}^{\dagger}\widehat{E}_{m}\widehat{C}\left|  \alpha\right\rangle$ is an effective Hamiltonian which is determined by the error-inducing Hamiltonians transformed by the coding and decoding and projected onto the initial state of the ancilla. From Eq.(\ref{matcod}) one can see that $\widehat{h}_{e}=0$ and hence $\widehat{\rho}_{I}$\ remains
constant in time: as long as we repeat the coding-decoding-ancilla resetting sequence, the information initially stored in $\mathcal{I}$\ is protected.

It is not always feasible to directly measure the ancilla independently from the information system ; in other words, it is sometimes impossible to perform a projection onto disentangled subspaces of $\mathcal{H}$\ of the form $\mathcal{H}_{I}\mathcal{\otimes}Span\left[  \left|  \alpha\right\rangle \right]  $ : in some cases, one can only project onto entangled subspaces of the total Hilbert space $\mathcal{H}$. In such a case the information initially stored
in the vector $\left|  \psi_{I}\right\rangle =\sum_{i=1}^{I}c_{i}\left|
\nu_{i}\right\rangle \in\mathcal{H}_{I}$ is transferred into an entangled state of $\mathcal{I}$ and $\mathcal{A}$ of the form $\left|  \psi \right\rangle =$ $\sum_{i=1}^{I}c_{i}\left|  \gamma_{i}\right\rangle $ where
the $I$ vectors $\left|  \gamma_{i}\right\rangle $ ($i=1,...,I$) which form an orthonormal basis of the information-carrying subspace $\mathcal{C}$, are not factorized as earlier but are in general entangled states. Nevertheless the same method as before can be used in that case to protect information, albeit in a different subspace $\mathcal{C}$.

To conclude this description of our method, let us now return to conditions (\ref{bascod1}) and (\ref{bascod2})\ imposed on the codewords $\left\{
\left|  \widetilde{\gamma}_{i}\right\rangle ,i=1,...,I\right\}  $ and make two remarks about them:

A. We can establish a useful relation between the dimension $A$ of the ancilla and
the number $M$ of correctable error Hamiltonians. The set of the $I$
codewords can indeed be seen as a collection of $2I\times N=2I^{2}A$ real numbers on
which $2I^{2}+2MI^{2}=2I^{2}(1+M)$ constraints, directly derived from
Eqs.(\ref{bascod1},\ref{bascod2}), are imposed. As the number of free parameters must be larger than the number of constraints, we necessarily have
$2I^{2}A\geq2I^{2}(1+M)$, or equivalently
\begin{equation}
A-1\geq M. \label{Hamming}
\end{equation}
This condition, called the ''Hamming bound'', gives an upper-bound on the number of independent error-inducing Hamiltonians that our method can correct simultaneously.

B. We may compare our correctability conditions (\ref{bascod2}) with the more general conditions of standard quantum
error-correction \cite{NC01, KL97}
\begin{align}
\forall\left(  \left|  \widetilde{\gamma}_{s}\right\rangle ,\left|
\widetilde{\gamma}_{t}\right\rangle \right)   &  \in\widetilde{\mathcal{C}
}^{2},\text{ }\forall\left(  \widehat{\mathbf{E}}_{k},\widehat{\mathbf{E}}%
_{l}\right)  \in\left\{  \widehat{\mathbf{E}}_{j}\left(  \left\{  \widehat
{E}_{m}\right\}  \right)  \right\}  ,\nonumber\\
\left\langle \widetilde{\gamma}_{t}\left|  \widehat{\mathbf{E}}_{k}^{\dagger
}\widehat{\mathbf{E}}_{l}\right|  \widetilde{\gamma}_{s}\right\rangle  &
=\alpha_{kl}\left\langle \widetilde{\gamma}_{t}|\widetilde{\gamma}%
_{s}\right\rangle \label{correcgene}%
\end{align}
which ensure the existence of a code space that is completely protected
against the error-inducing Hamiltonians $\widehat{E}_{m}$. Here $\alpha_{kl}$
are complex numbers, and the set $\left\{  \widehat{E}_{m}\right\}  $\ of
Hermitian operators $\widehat{E}_{m}$ generates a group $\mathcal{G}\left(
\left\{  \widehat{E}_{m}\right\}  \right)  $\ of all possible error-induced
evolutions (\ref{evolinf}). By $\left\{  \widehat{\mathbf{E}}_{j}\left(\left\{  \widehat{E}_{m}\right\}\right)  \right\}$ we denote a complete
basis set of operators which spans the space of evolution operators
$\widehat{U}$ and allows one to represent any $\widehat{U}$ as a linear combination of the basis operators $\widehat{\mathbf{E}}_{j}$. In addition to all the $\widehat{E}_{m}$, the variety of all linear combinations of $\widehat{\mathbf{E}}_{j}$ includes also many other operators given by commutators of all orders in $\widehat{E}_{m}$ entering the expansion of $\widehat{U}$ for long
times. The condition (\ref{correcgene}) is therefore much more restrictive
than Eq.(\ref{bascod2}). Moreover, even for two generic matrices
$\widehat{E}_{m}$, the basis $\left\{\widehat{\mathbf{E}}_{j}\right\}
$\ spans the entire Hilbert space $\mathcal{H}$, yielding $\widetilde
{\mathcal{C}}=\emptyset$. Only if the set $\left\{  \widehat{E}_{m}\right\}
$\ belongs to an extraspecial algebra restricting the error evolution
operators $\widehat{U}$ to a subgroup $\mathcal{G}\left(  \left\{  \widehat
{E}_{m}\right\}  \right)  \subset\mathcal{G}_{U}\left(  \mathcal{H}\right)  $
of the full unitary group in $\mathcal{H}$, a non-trivial code space
$\widetilde{\mathcal{C}}$\ may exist. The Zeno effect is the only way to
suppress loss of coherence if it is not the case.

\subsection{The code space and the coding matrix\label{III}}

It is sometimes possible to build the code space $\widetilde{\mathcal{C}}$\ explicitly from physical considerations. However, in general, we need an algorithm to calculate the code basis $\left\{  \left|
\widetilde{\gamma}_{i}\right\rangle \right\}  _{i=1,...,I}$ or, equivalently,
the coding matrix $\widehat{C}$. In this paragraph, we shall first describe this algorithm, then, we shall show that the non-holonomic control
technique can be employed to implement the coding matrix physically. We will also provide an algorithm which achieves the appropriate control.

Let us first make a remark which will be useful. Consider a
vector $\left|  \mathsf{C}\right\rangle $ of some Hilbert space and a matrix
$\widehat{\mathsf{E}}$ on this space. From the vector $\left|  \mathsf{C}\right\rangle $ we want to calculate a vector $\left|  \widetilde{\mathsf{C}}\right\rangle $ such that $\left\langle \widetilde{\mathsf{C}}\left|
\widehat{\mathsf{E}}\right|  \widetilde{\mathsf{C}}\right\rangle =0$. If
$\left\langle \mathsf{C}\left|  \widehat{\mathsf{E}}\right|  \mathsf{C}\right\rangle =0$, then $\left|  \mathsf{C}\right\rangle =\left|
\widetilde{\mathsf{C}}\right\rangle $ and the function $f_{\widetilde{\mathsf{C}}}(\lambda)=\left\|  \left|  \widetilde{\mathsf{C}}\right\rangle +\lambda\widehat{\mathsf{E}}\left|  \widetilde{\mathsf{C}}\right\rangle \right\|  ^{2},$ depending on the c-number $\lambda$, is minimal for $\lambda=0$: indeed
\begin{align*}
\left\|  \left|  \widetilde{\mathsf{C}}\right\rangle +\lambda\widehat
{\mathsf{E}}\left|  \widetilde{\mathsf{C}}\right\rangle \right\|  ^{2}  &
=\left\langle \widetilde{\mathsf{C}}|\widetilde{\mathsf{C}}\right\rangle
+\lambda\left\langle \widetilde{\mathsf{C}}\left|  \widehat{\mathsf{E}%
}\right|  \widetilde{\mathsf{C}}\right\rangle +\lambda^{\ast}\left\langle
\widetilde{\mathsf{C}}\left|  \widehat{\mathsf{E}}^{\dagger}\right|
\widetilde{\mathsf{C}}\right\rangle +\left|  \lambda\right|  ^{2}\left\langle
\widetilde{\mathsf{C}}\left|  \widehat{\mathsf{E}}^{\dagger}\widehat
{\mathsf{E}}\right|  \widetilde{\mathsf{C}}\right\rangle \\
& =1+\left|  \lambda\right|  ^{2}\left\langle \widetilde{\mathsf{C}}\left|
\widehat{\mathsf{E}}^{\dagger}\widehat{\mathsf{E}}\right|  \widetilde
{\mathsf{C}}\right\rangle ,
\end{align*}
and as $\left\langle \widetilde{\mathsf{C}}\left|  \widehat{\mathsf{E}}^{\dagger}\widehat{\mathsf{E}}\right|  \widetilde{\mathsf{C}}\right\rangle
\geq0$, $f_{\widetilde{\mathsf{C}}}(\lambda)$ is minimal for $\left|
\lambda\right|  =0$, that is $\lambda=0$. But, if $\left\langle \mathsf{C}\left|  \widehat{\mathsf{E}}\right|  \mathsf{C}\right\rangle \neq0$, we can
apply the following iterative method: we minimize $f_{\mathsf{C}}(\lambda)$\ with respect to $\lambda$, then we set $\left|  \mathsf{C}^{\prime
}\right\rangle =\left|  \mathsf{C}\right\rangle +\frac{\lambda}{2}%
\widehat{\mathsf{E}}\left|  \mathsf{C}\right\rangle $ and take $\frac{\left|
\mathsf{C}^{\prime}\right\rangle }{\sqrt{\left\langle \mathsf{C}^{\prime
}|\mathsf{C}^{\prime}\right\rangle }}$ as our new $\left|  \mathsf{C}%
\right\rangle $ ; repeating this sequence finally leads $\left|  \widetilde{\mathsf{C}}\right\rangle $, such that $\left\langle \widetilde{\mathsf{C}}\left|
\widehat{\mathsf{E}}\right|  \widetilde{\mathsf{C}}\right\rangle =0$.

Let us now return to our problem and show how to use the previous remark. We want to find $I$ vectors $\left|  \widetilde{\gamma}_{i}%
\right\rangle $\ which meet the conditions (\ref{bascod1}) and (\ref{bascod2})
; equivalently, we look for an orthonormal basis in which all
the matrices $\widehat{\mathsf{E}}_{k}$ have their $I \times I$ upper left
blocks equal to zero. To solve this problem, we propose to transform our initial problem in such a way that it can be
dealt with by the iterative algorithm presented in the previous paragraph. Let us combine the $I$ vectors $\left|  \widetilde{\gamma}_{i}\right\rangle $ into
a $\left(  N\times I\right)  $ ''supervector''
\[
\left|  \widetilde{\mathsf{C}}\right\rangle =\left(
\begin{array}[c]{c}
\left|  \widetilde{\gamma}_{1}\right\rangle \\
\vdots\\
\left|  \widetilde{\gamma}_{I}\right\rangle
\end{array}
\right)  .
\]
Then let us build $E=\left(  \frac{I(I-1)}{2}+M\frac{I(I+1)}{2}\right)  $
different $\left(  N\times I\right)  \times\left(  N\times I\right)
$-dimensional super-matrices $\widehat{\mathsf{E}}_{k}$ in the following way:
we consider them as made of $I^{2}$ blocks of dimension $N\times N$ and we successively fill each of these blocks with the different Hamiltonians
$\widehat{E}_{m}$ or the identity matrix $\widehat{I}$ or $0$. To be more
explicit, the first $\frac{I(I-1)}{2}$ matrices are built by simply placing
the $N\times N$ identity matrix in each of the $\frac{I(I-1)}{2}$ blocks
situated above the diagonal. In the last $\frac{MI(I+1)}{2}$ ones, the $M$
operators $\widehat{E}_{m}$ are successively placed in each of the
$\frac{I(I+1)}{2}$ blocks on and above the diagonal. One can thus reformulate
the conditions (\ref{bascod1}) as follows: for $1\leq k\leq\frac{I(I-1)}{2}$, $\left\langle \widetilde{\mathsf{C}}\left|  \widehat{\mathsf{E}}_{k}\right|
\widetilde{\mathsf{C}}\right\rangle =0.$
This form does not take the normalization condition into account,
which will be imposed in a different manner. Similarly, the conditions (\ref{bascod2}) are translated into the following form: for $\frac{I(I-1)}{2}+1\leq k\leq
\frac{I(I-1)}{2}+\frac{MI(I+1)}{2}$,
$\left\langle \widetilde{\mathsf{C}}\left|  \widehat{\mathsf{E}}_{k}\right|
\widetilde{\mathsf{C}}\right\rangle =0.$
This new problem can be handled by the same kind of iterative algorithm as in our preliminary remark. 

First, we randomly pick a supervector $\left|  \mathsf{C}_{0}\right\rangle $ which will be the starting point of the first step: we normalize this vector by imposing to each
of its $I$ components to have norm = $\frac{1}{I}$. If one of the components
of $\left|  \mathsf{C}_{0}\right\rangle $ is non normalizable, that is equals
zero, we pick up a new random supervector $\left|  \mathsf{C}_{0}\right\rangle
$ as a starting point.

Then, we minimize $F_{\mathsf{C}_{0}}\left(  \lambda_{1}^{(0)},\lambda_{2}^{(0)},...,\lambda
_{E}^{(0)}\right)  =\sum_{k=1}^{E}\left\|  \left|  \mathsf{C}_{0}\right\rangle
+\lambda_{k}^{(0)}\widehat{\mathsf{E}}_{k}\left|  \mathsf{C}_{0}\right\rangle
\right\|  ^{2}$ with respect to the $E$ c-numbers $\lambda_{k}^{(0)}$, and we calculate $\left|  \Delta
\mathsf{C}_{0}\right\rangle =\sum_{k}\lambda_{k}^{(0)}\widehat{\mathsf{E}}%
_{k}\left|  \mathsf{C}_{0}\right\rangle $ and $\left|  \mathsf{C}_{0}^{\prime
}\right\rangle =\left|  \mathsf{C}_{0}\right\rangle +\frac{1}{2}\left|
\Delta\mathsf{C}_{0}\right\rangle $. We normalize $\left|  \mathsf{C}%
_{0}^{\prime}\right\rangle $ by requiring each of its $I$ components to have
the norm = $\frac{1}{I}$, and take the result of this operation as our new
starting point $\left|  \mathsf{C}_{1}\right\rangle $. If one of the
components of $\left|  \mathsf{C}_{0}^{\prime}\right\rangle $ is non
normalizable, that is equals zero, we pick up a new random supervector
$\left|  \mathsf{C}_{0}\right\rangle $ as a starting point.

We repeat this sequence of operations as long as needed and obtain the desired vector $\left|  \widetilde{\mathsf{C}}\right\rangle$ asymptotically. Practically, as our algorithm converges quickly, the number of iterations needed is small.

\bigskip

\bigskip

The coding matrix $\widehat{C}$ is a complex unitary operator on the Hilbert space of the compound system
$\mathcal{S=I}\otimes\mathcal{A}$. We have just shown how to calculate the
codewords, which actually form the first $I$ columns of $\widehat{C}$, but one can wonder how to implement it physically. This question can be solved by the non-holonomic control technique.

Indeed, we can directly apply the results of the first of our articles to our coding problem in the following way: first, we find the codewords $\left\{\left|  \widetilde{\gamma}_{i}\right\rangle ,i=1,...,I\right\}  $ by the iterative algorithm we have previously presented, then we complete the set of $I$ vectors $\left\{  \left|  \widetilde{\gamma}_{i}\right\rangle ,i=1,...,I\right\}$ with $\left(  N-I\right)  $ vectors
$\left\{  \left|  \widetilde{\gamma}_{j}\right\rangle ,j=I+1,...,N\right\}  $
to form an orthonormal basis of $\mathcal{H}$, we build the coding matrix by
taking the vectors $\left\{  \left|  \widetilde{\gamma}_{i}\right\rangle
,i=1,...,N\right\}  $ as columns of $\widehat{C}$, and finally we calculate
the $N^{2}$ appropriate timings $\left\{  \tau_{i}\right\}  $ such that
\[
\widehat{U}\left(  \tau_{1},...,\tau_{N^{2}}\right)     =\exp\left(
-i\widehat{H}_{a}\tau_{N^{2}}\right)\ldots\exp\left(  -i\widehat{H}_{b}\tau_{1}\right)    =\widehat{C}
\]
through the complete control algorithm we have previously presented (we suppose we have two distinct perturbations $\widehat{P}_{a}$ and $\widehat{P}_{b}$ such that the system is completely controllable). Note that we assume $\widehat{H}_{0}=0$, hence $\widehat{H}_{a}=\widehat{P}_{a}$ and $\widehat
{H}_{b}=\widehat{P}_{b}$.

Actually, this straightforward procedure provides a lot of useless work. Indeed, most of the information contained in the coding matrix is irrelevant and the $N^{2}$ real
parameters of $\widehat{C}$\ do not all have to be controlled exactly: the
number $n_{C}$\ of necessary control parameters $\left\{  \tau_{i}\right\}  $\ is actually much less than $N^{2}$, as we shall see now.

The coding matrix is characterized by the relations (\ref{matcodbis}). The
problem of control thus reduces to finding $n_{C}$\ timings $\tau_{i}$, forming the time-vector $\overrightarrow{\tau}=\left(
\begin{array}
[c]{c}%
\tau_{1}\\
\vdots\\
\tau_{n_{C}}%
\end{array}
\right)  $, such that the non-holonomic evolution matrix
\[
\widehat{U}\left(  \overrightarrow{\tau}\right)  =\exp\left(  -i\widehat{H}_{a}\tau_{n_{C}}\right) \ldots\exp\left(  -i\widehat{H}_{a}\tau_{1}\right)
\]
checks (\ref{matcodbis}). The number $n_{C}$ of control parameters must exceed the number of independent constraints which is clearly $\sim MI^{2}$, that is $n_{C}\gtrsim MI^{2}$. Thus the
number of necessary control parameters appears to be much smaller than $N^{2}$. So we need a new algorithm which achieves a partial and less expensive control of the evolution operator of the system.

The algorithm we shall use to calculate the appropriate control timings $\tau_{i}$ mixes the iterative algorithm presented at the beginning of this paragraph and the non-holonomic control technique. If we introduce the $\left(N\times I\right)  \times\left(  N\times I\right)  $-dimensional block-diagonal
matrix
\[
\widehat{\mathsf{U}}\left(  \overrightarrow{\tau}\right)  =\left(
\begin{array}
[c]{cccc}%
\widehat{U}\left(  \overrightarrow{\tau}\right)  & 0 & \cdots & 0\\
0 & \widehat{U}\left(  \overrightarrow{\tau}\right)  & \cdots & 0\\
\vdots & \vdots & \vdots & \vdots\\
0 & 0 & \cdots & \widehat{U}\left(  \overrightarrow{\tau}\right)
\end{array}
\right)
\]
and the $\left(  N\times I\right)  $-dimensional supervector
\[
\left|  \mathsf{C}\right\rangle =\left(
\begin{array}
[c]{c}%
\left|  \gamma_{1}\right\rangle \\
\vdots\\
\left|  \gamma_{I}\right\rangle
\end{array}
\right)
\]
composed of the coordinates of the $I$ basis vectors of $\mathcal{C}$, we can
set our problem of control into the following equivalent
form: we look for a time-vector $\overrightarrow{\tau}$ such that
\begin{equation}
\forall k,\text{ \ }\left\langle \mathsf{C}\left|  \widehat{\mathsf{U}%
}^{\dagger}\left(  \overrightarrow{\tau}\right)  \widehat{\mathsf{E}}_{k}%
\widehat{\mathsf{U}}\left(  \overrightarrow{\tau}\right)  \right|  \mathsf{C}%
\right\rangle =0 \label{matcodquattro}%
\end{equation}
where the matrices $\left\{  \widehat{\mathsf{E}}_{k}\right\}  _{k=1,...,E}%
$\ denote $E$ different matrices of dimension $\left(  N\times I\right)
\times\left(  N\times I\right)$ which have been introduced in the beginning
of this section. The idea of our algorithm is to take the super
vector $\left|  \mathsf{C}_{0}\right\rangle =\widehat{\mathsf{U}}\left(
\overrightarrow{\tau}_{0}\right)  \left|  \mathsf{C}\right\rangle $, where
$\overrightarrow{\tau}_{0}$ is a random time-vector, as the starting point for an
elementary step of the iterative algorithm and look for the small time
increment $\overrightarrow{d\tau}_{0}$ such that $\widehat{\mathsf{U}}\left(
\overrightarrow{\tau}_{0}+\overrightarrow{d\tau}_{0}\right)  \left|  \mathsf{C}%
\right\rangle $\ follows the direction provided by the result $\left|
\mathsf{C}_{0}\right\rangle +\left|  \Delta\mathsf{C}_{0}\right\rangle $ of
the iterative algorithm. The repetition of this sequence finally yields
$\overrightarrow{\tau}=\overrightarrow{\tau}_{0}+\overrightarrow{d\tau}_{0}%
+\overrightarrow{d\tau}_{1}+...$ which meets Eq.(\ref{matcodquattro}).

Let us now describe the algorithm in more detail. First, we randomly pick a
set of timings $\tau_{0,i}$ in a ''realistic range'', dictated by the system
under consideration: in particular, control-pulse timings have to be much
shorter than the typical lifetime of the system but much longer than the
typical response delay required by the experiment. Then we calculate $\left|  \Delta
\mathsf{C}_{0}\right\rangle =\sum_{k}\lambda_{k}\widehat{\mathsf{E}}_{k}\left|  \mathsf{C}_{0}\right\rangle $ by minimizing the same function $F_{\mathsf{C}_{0}}\left(  \lambda_{1}^{\left(  0\right)  },\lambda
_{2}^{\left(  0\right)  },...,\lambda_{E}^{\left(  0\right)  }\right)$ as in the algorithm presented at the beginning of this section. At that point, we look for the
small increment $\overrightarrow{d\tau}_{0}$ of the time-vector $\overrightarrow
{\tau}_{0}$ such that
\begin{align}
\forall k,\left\langle \mathsf{C}\left|  \left(  \frac{\partial\widehat
{\mathsf{U}}^{\dagger}}{\partial\overrightarrow{\tau}}\left(  \overrightarrow
{\tau}_{0}\right)  .\overrightarrow{d\tau}_{0}\right)  \widehat{\mathsf{E}}%
_{k}\widehat{\mathsf{U}}\left(  \overrightarrow{\tau}_{0}\right)  +\widehat
{\mathsf{U}}^{\dagger}\left(  \overrightarrow{\tau}_{0}\right)  \widehat
{\mathsf{E}}_{k}\left(  \frac{\partial\widehat{\mathsf{U}}}{\partial
\overrightarrow{\tau}}\left(  \overrightarrow{\tau}_{0}\right)  ..\overrightarrow
{d\tau}_{0}\right)  \right|  \mathsf{C}\right\rangle \nonumber\\
=\frac{\left\langle \mathsf{C}_{0}+\frac{1}{2}\Delta\mathsf{C}_{0}\left|
\widehat{\mathsf{E}}_{k}\right|  \mathsf{C}_{0}+\frac{1}{2}\Delta
\mathsf{C}_{0}\right\rangle -\left\langle \mathsf{C}_{0}\left|  \widehat
{\mathsf{E}}_{k}\right|  \mathsf{C}_{0}\right\rangle }{\left\langle
\mathsf{C}_{0}+\frac{1}{2}\Delta\mathsf{C}_{0}|\mathsf{C}_{0}+\frac{1}%
{2}\Delta\mathsf{C}_{0}\right\rangle }. \label{equation}%
\end{align}

It should be noticed that we do not consider the error super-matrices
$\widehat{\mathsf{E}}_{k}$ corresponding to orthonormality conditions: in
other words, we just take matrices $\left\{  \widehat{\mathsf{E}}_{k}\right\}
_{k\in\left[  \frac{I(I-1)}{2}+1,\frac{I(I-1)}{2}+\frac{MI(I+1)}{2}\right]  }$
into account. Thus we deal with $\frac{MI(I+1)}{2}$ complex equations. This
set of equations can be reduced to the real linear system
\begin{equation}
\widehat{S}\left(  \overrightarrow{\tau}_{0}\right)  \cdot\overrightarrow{d\tau}%
_{0}=\overrightarrow{W}\left(  \left|  \Delta\mathsf{C}_{0}\right\rangle
\right)  \label{systeme}%
\end{equation}
where $\widehat{S}\left(  \overrightarrow{\tau}_{0}\right)  $ and
$\overrightarrow{W}\left(  \left|  \Delta\mathsf{C}_{0}\right\rangle \right)
$ are respectively an $MI^2\times n_{C}$ real matrix and a $MI^2$-dimensional real vector. We obtained Eq.(\ref{systeme}) by splitting the set
of $\frac{MI(I+1)}{2} $ complex equations (\ref{equation}) into two sets of
$\frac{MI(I+1)}{2}$ real equations, and rejecting those which are trivial
($0=0$) or redundant. Though straightforward, the
explicit expressions of the different elements of $\widehat{S}$ and
$\overrightarrow{W}$ involve many indices and are so unpleasant that we prefer not to reproduce them here.

The linear system we have just found is, a priori, rectangular $(MI^2\times n_{C})$, but actually the number $n_{C}$ has not been fixed yet. Previously, we stated that $n_{C}\geq MI^2$: we could be tempted to set $n_{C}=MI^2$ so as to obtain a square system, easily solvable by standard techniques of linear algebra. Yet we will proceed in a slightly different way. We set $n_{C}=MI^2+\delta n>MI^2$, where $\delta n$ is an integer of order $1$, then we randomly pick $MI^2$ timings $t_{i}$ among the $n_{C}$ which will be considered as free parameters, whereas the other $\delta n$ ones will be regarded as frozen. The new version of Eqs(\ref{systeme}) is now clearly a square system, which yields the $MI^2$-dimensional increment $\overrightarrow{d\tau}_{0}$, corresponding to the $MI^{2}$ free varying timings, which we complete
with $\delta n$ zeros, corresponding to the frozen timings, into a $n_{C}$-dimensional vector $\overrightarrow{d\tau}_{0}$. Then we set $\overrightarrow{\tau}_{1}^{\alpha}$ $=\overrightarrow{\tau}_{0}+$ $\alpha$ $\overrightarrow{d\tau}_{0}$
where $\alpha$ is a convergence coefficient and calculate the test function
$G\left(  \overrightarrow{t}\right)  =\sum_{k}\left|  \left\langle
\mathsf{C}\left|  \widehat{\mathsf{U}}^{\dagger}\left(  \overrightarrow
{\tau}\right)  \widehat{\mathsf{E}}_{k}\widehat{\mathsf{U}}\left(
\overrightarrow{\tau}\right)  \right|  \mathsf{C}\right\rangle \right|  ^{2}$ in
$\overrightarrow{\tau}=\overrightarrow{\tau}_{1}^{\alpha}$ for different values of
$\alpha\in\left[  0,1\right]  $. If we find an $\alpha_{1}$ such that
$G\left(  \overrightarrow{\tau}_{1}^{\alpha}\right)  <G\left(  \overrightarrow
{\tau}_{0}\right)  $, we take $\overrightarrow{\tau}_{1}\equiv\overrightarrow{\tau}_{1}^{\alpha_{1}}$ as our new time-vector, and keep the same free-varying
timings. If we cannot find such an $\alpha_{1}$, this means we are situated in a local minimum of $G$ ; then we set $\overrightarrow{\tau}_{1}\equiv\overrightarrow{\tau}_{0}$ and pick
a new set of free varying parameters. This rotation procedure among control
parameters allows us to avoid possible local minima of the test function $G$ we want to cancel.

We repeat this sequence of operations as long as needed and obtain the desired vector $ \overrightarrow{\tau}$ asymptotically. Practically, as our algorithm converges quickly, the number of iterations needed is small.

\bigskip

We have not said anything about decoding so far. If the signs of the two Hamiltonians
$\widehat{H}_{a}=\widehat{P}_{a}$ and $\widehat{H}_{b}=\widehat{P}_{b}$ can be
reversed by altering the
control field parameters, decoding amounts to reversing $\widehat{P}_{a}$ and $\widehat{P}_{b}$ and
applying the same control timing sequence backwards. Otherwise, one must use the general
non-holonomic control technique, involving $N^{2}$ control parameters, to find
timings which realize $\widehat{C}^{-1}$.

\section{ Zeno Coherence Protection by Random Coding}

The protection method we presented in the previous section seems promising for relatively low-dimensional systems. However, for large systems, it is
likely to lead to very heavy computations and long control sequences. To deal with such systems, we therefore propose
to employ an approach inspired by classical random coding \cite{SW40} : in this method, linear codes $[n,k,d]$ are produced, in which $k$-bit words are encoded as randomly chosen
$n$-bit sequences. The minimal Hamming distance $d$ between any two codewords
approaches the Hamming bound $d\leq n-k$ as $n\rightarrow\infty$. In this
section, we show how to extend the idea of random coding to the quantum case.

Strong mixing or entanglement occur in the phase or Hilbert spaces of complex classical or quantum systems, respectively, and can, in principle, be used for random coding. However, in practice, in the classical case, the dynamics of such systems is not reversible, which makes subsequent decoding hardly possible. By contrast, the dynamics of multi-dimensional quantum systems can be reversed, when the underlying physical mechanism is simple enough : the spin-echo phenomenon is a typical example of this topic. High dimensionality of simple quantum systems is responsible for the massive parallel computing capacity of quantum computers. Therefore, we have to find an operation which produces strong mixing in the multidimensional Hilbert space, and which can be inverted in a simple way : the non-holonomic control suits perfectly this purpose. The essential requirement for the protection scheme we propose to apply is that error-inducing interactions are relatively simple, resulting, for instance, either from a binary qubit interaction or, generally speaking, from a few-particle coupling.

To combine strong mixing with irreversibility, we assume that we have a quantum system with a large number of separate energy levels and with two simple interactions which satisfy the bracket generation condition and
can therefore be employed for the non-holonomic control. In such a system, one can encode quantum data into strongly mixed states by straightforwardly applying a unitary transformation $\widehat{C}$, the decoding procedure being achieved by the inverse transformation $\widehat{C}^{-1}$. Encoding the data into many levels
allows us to strongly reduce the error-rate, as will be shown. In turn, by applying the Zeno effect \cite{MS77, KK00} we can restore the slightly corrupted data back to its original value with high probability.

To encode the data in a high dimensional Hilbert space of $n$ qubits, we
introduce $n-k$ ancillary qubits in addition to the required number $k$ of data
qubits. This results in an increase in the number of possible errors, which
depends polynomially on $n$, but the error rate decreases at will because the
infinitesimal errors are semi-orthogonal to the encoded data to a degree
exponential in $n-k$. The degree of semi-orthogonality reflects the error
correction efficiency of the coding. Efficiency requires a precise and careful choice of
the code for coding in minimal dimensions, but in high dimensions it is
naturally achieved by random coding. In mathematical terms, the method relies
on the fact, that in a multidimensional space, a pair of randomly chosen
vectors are almost orthogonal with high probability. In physical terms, random
coding is equivalent to strong mixing, or full population of all energy levels,
which can be reached by the non-holonomic control with a number of interaction
switchings depending only polynomially on $n$. Thus the random coding approach
of the present section complements our earlier non-holonomic Zeno coherence loss
suppression scheme, which requires exponential effort to find and achieve an
exact code, and is thus efficient only for low-dimensional systems.

The essence of random coding can be elucidated as follows. Consider an $n$-qubit system, comprising a $k$-qubit information carrying subsystem and an $(n-k)$-qubit ancilla. In the $2^{n}$-dimensional Hilbert space of the system, the error-inducing Hamiltonians $\widehat{E}_{m}$ corresponding to a few-particle interaction are represented by sparse matrices in the computational basis, which is composed of all the possible tensor products of individual qubits eigenstates. In this basis, the number of non-zero matrix elements is indeed polynomial in $n$ : for instance, for binary interactions, this number scales as $n^{2}$ . The coding-decoding transformation $\widehat{C}^{\dagger}\widehat{E}_{m}\widehat{C}$, where $\widehat{C}$ stands for a generic unitary matrix, 'smoothes' all the matrix elements by mixing them : finally, all these elements are of the same order of magnitude, which is, up to a polynomial factor, $2^{n}$-times smaller than the typical value of non-zero matrix elements in the computational basis before the coding-decoding sequence. Error matrices elements are thus exponentially reduced : the error suppression condition Eq.(\ref{matcodbis}) for the projection of these matrices onto the subspace of the initial state of the ancilla is not fulfilled exactly any longer : the projection
differs from zero, but its norm $\left\langle \gamma_{t}\left\vert \widehat{C}^{\dagger}\widehat{E}_{m}%
\widehat{C}\right\vert \gamma_{s}\right\rangle \sim2^{k-n}$
remains small, and decreases exponentially with the size of the ancilla. The error-accumulation rate is thus inhibited by a factor of the order of the ancilla Hilbert space dimension.

Note, that this mechanism is efficient only when the generic coding matrix
$\widehat{C}$ can be achieved by a small number of switchings, such that
the coding procedure does not take exponentially long time. Fortunately, the coding matrix takes a generic form after a relatively small number of switchings, which scales linearly with the number of qubits $n$ (see \cite{Gershkovich:00}). Moreover, if the signs of the two interactions employed for the
non-holonomic control can be inverted, the decoding operation can be performed at the same level of complexity as the coding procedure, by straightforwardly changing the signs of the interactions and inverting the timing sequence in which these interactions are applied. The main restriction to practical implementation of the random coding protection scheme arises from that one necessarily has to remain in the Zeno regime: the measurement time, which does not depend on the size of the system, has to be much shorter than the coherence loss timescale, which decreases, although polynomially, with the size of the system.

\section{Conclusion}

The non-holonomic control allied with the Quantum Zeno Effect can be employed to overcome the influence of the environment on the quantum system considered. On the one hand, in the case of low-dimensional systems, we showed that quantum information can be protected by frequently repeating the cycle coding-infinitesimal errors-decoding-projection : coding and decoding correspond to a unitary transformation of the Hilbert space and its inverse, respectively, which are determined in such a way that the projection onto the initial information carrying subspace of the state resulting from coding-infinitesimal errors-decoding yields the initial state vector. All the needed algorithmic tools have been presented. On the other hand, for high-dimensional systems, one can adapt the classical idea of random coding to the quantum case : the basic principle is to use non-holonomic control to impose generic and easily reversible unitary evolutions to the system in order to encode/decode the information ; this procedure ''dilutes'' the influence of the errors in the large Hilbert space and then decreases their influence.


\begin{thebibliography}{99}                                                                                       


\bibitem {Sho95}P.W. Shor, Phys. Rev. A \textbf{52}, 2493 (1995).


\bibitem{MS77}B. Misra and E.C.G. Sudarshan, J. Math. Phys. \textbf{18}, 756 (1977).

\bibitem{KK00}A.G. Kofman and G. Kurizki, Nature (London) \textbf{405},
546 (2000). 

\bibitem{Brion04}E. Brion, G. Harel, N. Kebaili, V. M. Akulin and I. Dumer, Europhys. Lett. \textbf{66}, 157 (2004). 

\bibitem{NC01}M.A. Nielsen and I.L. Chuang, ''Quantum Computation and
Quantum Information'', Cambridge University Press, 2001.

\bibitem{KL97}E.Knill and R. Laflamme, Phys. Rev. A \textbf{55}, 900 (1997).


\bibitem{SW40}C.E. Shannon and W. Weaver, ''The mathematical theory of information'', University of Illinois Press, Urbana (1940).

\bibitem {Gershkovich:00}V.~Gershkovich \textit{et al.}, IHES preprint IHES/P/00/01.

\end{thebibliography}
\end{document}